\documentclass[a4paper]{article}

\usepackage{amssymb} 
\usepackage{amsmath} 
\usepackage{color}
\usepackage{ mathrsfs }

\usepackage[top=3 cm, bottom=3 cm, left=4 cm, right= 4 cm ]{geometry}

\title{\bf Sterile Neutrino Dark Matter and Low Scale Leptogenesis from a Charged Scalar}
\author{Michele Frigerio\\
{\it \small Laboratoire Charles Coulomb, UMR 5221 (CNRS/Universit\'e Montpellier 2)}\\
{\it \small F-34095 Montpellier, France}
\and
Carlos E. Yaguna\\{\it \small Institut f\"ur Theoretische Physik, Universit\"at M\"unster,}\\
{\it \small Wilhelm-Klemm-Stra\ss e 9, D-48149 M\"unster, Germany}
}
\date{}
\begin{document}
\maketitle

\begin{abstract}
\noindent
We show that novel paths to dark matter generation and baryogenesis
are open when the Standard Model  is extended with three
sterile neutrinos $N_i$ and a charged scalar $\delta^+$.
Specifically, we propose a new production mechanism for the dark matter particle -- a multi-keV sterile neutrino, 
$N_1$ -- that does not depend on the active-sterile mixing angle and does not
rely on a large primordial lepton asymmetry. Instead, $N_1$ is 
produced, via freeze-in, by the decays of $\delta^+$ while it is
in equilibrium in the early Universe. In addition,  we demonstrate that, thanks to the couplings between the heavier sterile neutrinos $N_{2,3}$ and $\delta^+$, baryogenesis via leptogenesis can be realized close to the electroweak scale. The lepton asymmetry is generated either by $N_{2,3}$-decays for masses $M_{2,3}\gtrsim$ TeV, 
or by $N_{2,3}$-oscillations for $M_{2,3}\sim$ GeV. Experimental signatures of this scenario include an X-ray line from dark matter decays,
and the direct production of $\delta^+$ at the LHC.
This model thus describes a minimal, testable scenario for neutrino masses, the baryon asymmetry, and dark matter.
\end{abstract}

\section{Introduction}

Sterile neutrinos, that is, fermions singlet under the $SU(3)\times SU(2)\times U(1)$ gauge symmetry, are a very well-motivated extension of the Standard Model (SM). 
On the theoretical side, they are a prediction of left-right symmetric theories, they allow to gauge $B-L$
by removing its anomaly and they are necessary in $SO(10)$ grand unification. On the phenomenological side, they provide
a non-vanishing mass to the active SM neutrinos, they allow to realize baryogenesis via leptogenesis, and they 
are also a viable candidate for dark matter, as long as their mass lies in the keV range.
In that case, they are naturally long-lived so that, unlike  dark matter candidates at the electroweak scale, 
no additional symmetries are required to stabilize them. 
Thus, sterile neutrinos  can provide a simple solution to the three open problems of the SM: neutrino masses, the baryon asymmetry, and the dark matter.

The minimal model addressing these three issues requires one sterile neutrino $N_1$ at the keV scale as dark matter candidate \cite{Dodelson:1993je}, and two additional sterile neutrinos $N_{2,3}$ for leptogenesis, which is induced either by $N$-decays, for sterile neutrino masses  above the TeV scale
\cite{Fukugita:1986hr}, or by $N$-oscillations, for sterile neutrino masses at the GeV scale \cite{Akhmedov:1998qx,Asaka:2005pn}.
Light active neutrino masses are easily generated, via the seesaw mechanism, provided the sterile neutrino masses are significantly larger than about $1$ eV. 
Thus, one may argue that no new physics is needed above the electroweak scale to explain neutrino masses, baryogenesis, and dark matter, defining the so-called
``$\nu$ Minimal Standard Model" ($\nu$MSM) \cite{Asaka:2005an}.
Such a model is undoubtedly economical 
and very predictive, by reason of the small number of degrees of freedom it contains, but it is also strongly constrained. 
Present experiments, in fact,  already exclude the minimal scenario for sterile neutrino dark matter within the $\nu$MSM \cite{Canetti:2012kh}. In the region of mass 
and mixing angle compatible with current observations, in particular the X-ray bounds \cite{Abazajian:2006yn,Boyarsky:2007ay,Horiuchi:2013noa} and the $\mathrm{Ly}_\alpha$ forest data \cite{Boyarsky:2008xj, Viel:2006kd,Viel:2013fqw}, one cannot produce a large enough population of sterile neutrinos  from standard active-sterile oscillations \cite{Dodelson:1993je}. 
The only  way out within the $\nu$MSM is to generate the dark matter neutrinos via resonant active-sterile oscillations triggered by large initial lepton asymmetries \cite{Shi:1998km,Laine:2008pg}, but that mechanism requires a high degree of fine-tuning \cite{Shaposhnikov:2008pf,Canetti:2012kh}.  
In addition, leptogenesis via $N_{2,3}$-oscillations within the $\nu$MSM also requires a significant tuning of parameters, in particular a strong mass degeneracy between
$N_2$ and $N_3$ \cite{Canetti:2012kh}. 

It is important, therefore, to consider alternative ways of realizing leptogenesis and producing sterile neutrinos   within extensions of the $\nu$MSM.  
 Indeed, most extensions of the SM, that are theoretically well-motivated, contain
new degrees of freedom beside the sterile neutrinos, which  may modify significantly the phenomenology of the $\nu$MSM.  For leptogenesis, the required tuning of parameters  can be released by introducing a second Higgs doublet, as recently studied in \cite{Shuve:2014zua}. Regarding dark matter,  several other possibilities have been considered for the production of the sterile neutrinos. 
They include the production through inflaton decay \cite{Shaposhnikov:2006xi,Bezrukov:2009yw,Bezrukov:2014nza},  thermal overproduction followed by entropy dilution \cite{Bezrukov:2009th}, 
and the production via the decays of either a  neutral 
scalar  in thermal equilibrium \cite{Kusenko:2006rh}, or a frozen-in neutral scalar \cite{Merle:2013wta}.

In this paper we advocate an extension of the  SM by three sterile neutrinos, $N_i$, plus a  charged scalar, $\delta^+$, both of which are naturally present in left-right symmetric or unified extensions of the SM \cite{Frigerio:2006gx}. The charged scalar interacts with SM leptons, both doublets and singlets, and with sterile neutrinos, and its mass can be as low as allowed by current collider bounds  -- about $200$ GeV. We will show that the interplay between $N_i$ and $\delta^+$ modifies the dark matter phenomenology, enables  new realizations of baryogenesis, and gives rise to novel experimental signatures at colliders.  The decays of $\delta^+$, in fact,  constitute a new mechanism for the production of sterile neutrino dark matter in the early Universe. An advantage of this mechanism is that the resulting  relic density  does not depend on the active-sterile mixing angle, allowing to satisfy the strong bounds derived from X-ray observations \cite{Abazajian:2006yn,Boyarsky:2007ay,Horiuchi:2013noa}.  Besides, 
the charged scalar  induces independent sources of leptogenesis that are effective at scales close to the electroweak scale. 
If the charged scalar mass were accessible at colliders, the present scenario for dark matter and baryogenesis would be testable in a direct way.

The rest of the paper is organized as follows. In section \ref{sec:mod} we introduce the model and fix our notation. Section \ref{DM} deals with dark matter production via the decays of $\delta^+$. We obtain, in particular, the dark matter relic density as a function of the parameters of the model.  Baryogenesis via leptogenesis is discussed in section \ref{barlep}. 
We examine two different ways to generate a lepton asymmetry, and outline the region of the parameter space that successfully realizes baryogenesis in each case. The most interesting experimental signatures of this model are discussed in section \ref{exp}. 
Section \ref{flavour} briefly elaborates on  the flavour structure of the masses and Yukawa couplings of the model.
We present our conclusions in section \ref{sec:con}.

\section{The model}
\label{sec:mod}

When gauge singlet chiral fermions $N_{Ri}$ are added to the SM, they have in general a Majorana mass term and a Yukawa coupling to the SM lepton doublets, 
\begin{equation}
\label{eq:seesaw}
\mathscr{L}_N = \overline {N_{Ri}}i\gamma^\mu \partial_\mu N_{Ri} + \left[
-\frac 12 \overline{(N_{Ri})^c} (M_N)_{ij} N_{Rj} -\overline{l_{L\alpha}} (y_\nu)_{\alpha i} N_{Ri} \tilde{H} +h.c.\right]~.
\end{equation}
When a scalar field $\delta^+$, singlet under $SU(2)$ and with electromagnetic charge one, is added to the SM, it has in general, besides its  gauge and self-interactions,
a quartic coupling to the Higgs doublet and an antisymmetric Yukawa coupling to two lepton doublets,
\begin{equation}\begin{array}{rcl}
\mathscr{L}_\delta &=& D_\mu\delta^+D^\mu\delta^- - M_\delta^2 \delta^+\delta^- - \dfrac 12 \lambda_\delta (\delta^+\delta^-)^2
- \lambda_{\delta H} \delta^+\delta^- H^\dagger H \\
&+& \left[-\overline{l_{L\alpha}} (y_L)_{\alpha\beta} (i\sigma_2) (l_{L\beta})^c \delta^+ + h.c. \right] ~.
\end{array}
\label{Ldelta}\end{equation}
When both the sterile neutrinos $N_{Ri}$ and the charged scalar $\delta^+$ are present, there is an additional Yukawa coupling 
involving the SM lepton singlets,
\begin{equation}
\label{eq:lint}
\mathscr{L}_{\delta N}=  - \overline{(e_{R\alpha})^c} (y_R)_{\alpha i} N_{Ri} \delta^+ + h.c. ~.
\end{equation}
These three  terms, equations (\ref{eq:seesaw})-(\ref{eq:lint}), plus the well-known SM ones,  constitute the Lagrangian  considered in this paper. It is the most general Lagrangian consistent with the $SU(3)\times SU(2)\times U(1)$ gauge symmetry and with the addition of the singlet fermions and the charged scalar. Without loss of generality, we adopt the basis where $M_N$ and the charged lepton mass matrix are real and diagonal. For our following discussion,  the parameters of greater relevance are the masses of the singlet fermions ($M_i$, $i=1,2,3$) and of the charged scalar ($M_\delta$), and the $3\times 3$ Yukawa matrices $y_\nu$ and $y_R$.

In this model active neutrino masses are not affected by the existence of $\delta^+$ and  are obtained via the usual seesaw mechanism: $m_\nu \simeq y_\nu M_N^{-1} y^T_\nu v^2$,
 with $v\simeq 174$ GeV. 
In section \ref{DM} we will identify the sterile neutrino $N_1$ with a multi-keV dark matter candidate, with very small mixing angles with active neutrinos, corresponding to tiny neutrino Yukawa couplings,  $(y_\nu)_{\alpha 1}\lesssim 3\times 10^{-13}$. These tiny couplings give
rise to a lightest neutrino mass $m_\nu^\mathrm{lightest}\lesssim 10^{-6}~\mathrm{eV}$. The much larger solar and atmospheric neutrino mass scales are generated instead by  the seesaw contribution of the two heavier sterile neutrinos, $N_{2,3}$.  The active neutrino mass spectrum is thus hierarchical (either normal or inverted), with one neutrino remaining essentially massless.

The generation of the dark matter energy density and of the baryon asymmetry have been extensively studied in the context of the SM extended with sterile neutrinos only \cite{Canetti:2012kh}.
The SM extension with sterile neutrinos and $\delta^+$ was considered
in \cite{Frigerio:2006gx}. It was shown that, taking the sterile neutrinos $N_i$ as light as a few TeVs and heavier than $\delta^+$, one can generate the observed baryon asymmetry via leptogenesis, 
with no need to enhance  resonantly the CP asymmetry. 

We explore, instead, the possibility that the lightest singlet fermion, $N_1$,  has a multi-keV mass, $M_1 \ll M_{\delta}$, and accounts for the observed dark matter density.
 In this framework, the  existence of $\delta^+$ offers an alternative way of producing dark matter that has not been studied before in the literature. Moreover, we will show that this different mass spectrum is still compatible with the leptogenesis mechanism proposed in  \cite{Frigerio:2006gx} and,
in addition, we will identify a different mechanism for leptogenesis in the presence of $\delta^+$ based on $N_{2,3}$-oscillations.

\section{Dark matter production  from $\delta^+$ decays \label{DM}}

In this model, the only particle that can play the role of  dark matter  is the lightest singlet fermion, $N_{1}$. It is important, therefore, to determine how it is  produced in the early Universe and whether it can account for the observed dark matter density while respecting all other experimental  constraints. In this section, after briefly reviewing the standard scenario for sterile neutrino dark matter, we propose a new mechanism for dark matter production via the decays of $\delta^+$. 

In the absence of $\delta^+$, our model is described by the seesaw Lagrangian, equation (\ref{eq:seesaw}), and the region of the parameter space compatible with dark matter is the one of
the $\nu$MSM. It features a singlet fermion ($N_1$) with a mass at the keV scale, the dark matter particle, and two heavier singlets, $N_{2,3}$. In the $\nu$MSM, dark matter is produced at temperatures of order $100$ MeV via active-sterile neutrino mixing. The required effective mixing angle, $\theta_1$, is necessarily small and is related to the Yukawa couplings by 
 $\theta^2_1=\sum_{\alpha=e,\mu,\tau} |(y_\nu)_{\alpha 1}|^2 v^2/M_1^2$. 
Detailed numerical studies have shown that the observed dark matter density can be reproduced in two different cases \cite{Canetti:2012kh}. In the non-resonant case (also known as the Dodelson-Widrow mechanism \cite{Dodelson:1993je}),  dark matter neutrinos are thermally and non-resonantly produced  with a smooth distribution of momenta. This mechanism yields the minimal amount of dark matter that can be obtained for a given mass and mixing angle. For dark matter masses between $1$ keV and $10$ keV, the mixing angle required to explain the dark matter density is  $\sin^2(2\theta_{1})\sim 10^{-8}\mbox{-}10^{-9}$, as shown e.g. in figure 2 of \cite{Canetti:2012kh}. These parameters  are not consistent with current data \cite{Asaka:2006nq}, which imply $M_1\lesssim 3-4~\mathrm{keV}$ from the X-ray line bound \cite{Abazajian:2006yn,Boyarsky:2007ay,Horiuchi:2013noa}  and $M_1\gtrsim 8~\mathrm{keV}$ from the $\mathrm{Ly_\alpha}$ forest observations \cite{Boyarsky:2008xj, Viel:2006kd,Viel:2013fqw}. The other case in which the dark matter density can be explained, and the only one that is currently viable in the $\nu$MSM, is resonant production (also known as the Shi-Fuller mechanism \cite{Shi:1998km}). In it, the dark matter production rate is resonantly amplified by the presence of a lepton chemical potential in the plasma, which enhances the production for particular momenta as they pass through the resonance, giving rise to a non-thermal momentum distribution that is \emph{colder} than that obtained in the non-resonant case. To explain the observed dark matter density, a lepton asymmetry $|\mu_\alpha|\gtrsim 8\times 10^{-6}$ at $T\sim 100~\mathrm{MeV}$ is required \cite{Laine:2008pg}, where $\mu_\alpha=n_\alpha/s$, $s$ is the entropy density of the Universe, and $n_\alpha$ is the total number density of active leptons of flavour $\alpha$. In principle, this large lepton asymmetry can be generated within the $\nu$MSM via CP-violating oscillations of $N_2$ and $N_3$, but only if their mass difference, $M_3-M_2$, is fine-tuned to the order of $10^{-11}$ \cite{Shaposhnikov:2008pf,Roy:2010xq}.

The existence of $\delta^+$  allows a new production mechanism for sterile neutrinos  in the early Universe. Indeed, thanks to the interaction term in equation (\ref{eq:lint}), the decays of the charged scalars can produce sterile neutrinos, $\delta^+\to N_1\ell^+$. This decay is a thermal process that takes place  while $\delta^+$  is in equilibrium and it is an example of the so-called freeze-in scenario \cite{Hall:2009bx} for dark matter production. The defining feature of freeze-in is that the dark matter interactions are so weak that they never reach thermal equilibrium in the early Universe. As a result,  the dark matter abundance is negligible at high temperatures and slowly increases as the Universe cools down. The production ceases when the freeze-in temperature is reached, and the dark matter abundance remains constant from then on.

The $N_1$ yield, $Y_{N_1}(T)=n_{N_1}(T)/s(T)$, from $\delta^+$ decays is obtained by solving the following Boltzmann equation \cite{Hall:2009bx} 
\begin{equation}
 sT\frac{dY_{N_1}}{dT}=-\frac{\gamma_{N_1}(T)}{H(T)}
\label{eq:yield}
\end{equation}
where $s$ is the entropy density, $H$ is the expansion rate and $\gamma_{N_1}(T)$ is the thermally-averaged production rate. We have that
\begin{equation}
 \gamma_{N_1}(T)=\frac{M_\delta^2 T}{2\pi^2} K_1(M_\delta/T)\,\sum_\alpha \left[\Gamma(\delta^-\to N_1 \ell^-_\alpha)+\Gamma(\delta^+\to N_1 \ell^+_\alpha)\right]
\end{equation}
where $K_1(x)$ is the Bessel function of the second kind and the sum runs over the different lepton flavours.  Since the lepton and $N_1$ masses are negligible compared to $M_\delta$, 
the $\delta^+$ decay rates  are calculated as
\begin{equation}
  \Gamma(\delta^-\to N_1 \ell^-_\alpha)=\Gamma(\delta^+\to N_1 \ell^+_\alpha)=\frac{M_{\delta}|(y_R)_{\alpha 1}|^2}{16\pi}
\label{eq:drate}
\end{equation}
and their sum over flavours can be conveniently written as
\begin{equation}
  \sum_\alpha \left[\Gamma(\delta^-\to N_1 \ell^-_\alpha)+\Gamma(\delta^+\to N_1 \ell^+_\alpha)\right] =\frac{ M_\delta}{8\pi} \left(y_R^\dagger y_R\right)_{11}= 
\frac{ M_\delta}{8\pi} y_{R1}^2~,
\end{equation}
where we have defined $y_{R1}$ as the combination of couplings that determines the production of $N_1$ via $\delta^+$ decays. 
Using $s(T)=2\pi^2g_sT^3/45$, $H(T)=1.66\sqrt{g_\rho}T^2/M_{Pl}$, and $x\equiv M_{\delta}/T$ we can solve equation (\ref{eq:yield}) as
\begin{equation}
Y_{N_1}(T_0)=\frac{45}{(1.66) 32\pi^5g_s\,\sqrt{g_\rho}}\frac{M_{Pl}  y_{R1}^2} {M_\delta}\int_{x_{min}}^{x_{max}}x^3\,K_1(x)\,dx\,,
\end{equation}
where  $T_0$ is the temperature today and we used $Y(T\gg M_\delta)=0$ as our initial condition. 
Integrating this equation from $x_{max}=\infty$ to $x_{min}=0$ and setting $g_s \approx g_\rho\approx 100$ yields
\begin{equation}
 Y_{N_1}(T_0)\approx 1.6\times 10^{-5}\left(\frac{ y_{R1}  }{10^{-8}}\right)^2\left(\frac{1~\mathrm{TeV}}{M_\delta}\right).
\end{equation}
The $N_1$ relic density is then
\begin{equation}
 \Omega_{N_1} h^2\approx 0.11 \left(\frac{M_{1}}{\mathrm{keV}}\right)\left(\frac{ y_{R1}  }{5\times 10^{-8}}\right)^2\left(\frac{1~\mathrm{TeV}}{M_\delta}\right).
\label{relic}
\end{equation}
Thus, a keV sterile neutrino can explain the observed dark matter density via $\delta^+$ decays if $M_\delta\sim 1~\mathrm{TeV}$ and $ y_{R1} \sim \mathrm{few}~\times 10^{-8}$.  
This analytical result is in  very good agreement with the numerical calculations  we did for similar models of freeze-in dark matter \cite{Klasen:2013ypa,Molinaro:2014lfa}.

A crucial feature of equation (\ref{relic}) is that, unlike the production in the $\nu$MSM, it does not depend on the active-sterile mixing angle
$\theta_1$ that determines the decay rate of the sterile neutrino. Hence, one can take $\theta_1$ small enough to be in agreement with the X-ray bounds without affecting the predicted dark matter density. In other words,  in this model one can decouple the  dark matter production (determined by the $y_R$ couplings) from the dark matter decay (determined by $y_\nu$ via mixing angles),  opening  new viable regions and alleviating the experimental constraints on sterile neutrino dark matter. 
In particular,  $M_1$ can be  larger than the keV scale, as long as the dark matter sterile neutrino is sufficiently long-lived. Since $\Gamma_{N_1}\propto M_1^5 \sin^2\theta_1$,  one cannot exclude, for example, the possibility of vanishing active-sterile mixing, $\theta_1=0$, which would allow masses as high as $1$ MeV. Above that value, the decay $N_1\to \nu e^+ e^-$ becomes kinematically available and   
can proceed through the couplings $y_R$ and $y_L$, even for $\theta_1=0$.  In that case, the strong constraints coming from indirect searches
of electrons and positrons come into play, but they compete with the smallness of $y_{R1}$, determined by equation (\ref{relic}), and of the Yukawa matrix $y_L$, that can
be very small too.  

Regarding structure formation, the sterile neutrinos produced via $\delta^+$ decays are \emph{colder} than those obtained in the $\nu$MSM  \cite{Petraki:2007gq,Kusenko:2009up} because the decays take place at a temperature $T\sim M_\delta$ much higher than the QCD scale ($\sim 150~\mathrm{MeV}$). In fact, their free-streaming length and phase space density are  identical to those of the  so-called chilled sterile  neutrinos studied in \cite{Petraki:2008ef}. Consequently, the Lyman-$\alpha$ bounds on the sterile neutrino mass in our scenario are significantly weaker, allowing for a dark matter mass $M_1$ smaller than in the $\nu$MSM \cite{Kusenko:2013saa}. 

We assume in the following that the decays of the charged scalars are the dominant source of sterile neutrinos, so that equation (\ref{relic}) is satisfied. This will be the case in the absence of a large lepton asymmetry and for small mixing angles, $\sin^22\theta_1<10^{-9}$. Next, we show that this assumption is consistent with baryogenesis, and that it leads to new experimental signatures.

\section{Baryogenesis via leptogenesis with $\delta^+$ \label{barlep}} 

In this section we will discuss leptogenesis in the presence of the charged scalar field $\delta^+$. 
We will briefly review the two basic mechanism for leptogenesis in the minimal scenario with sterile neutrinos only: 
$N$-decays and $N$-oscillations. We will
show that both mechanisms are still operative when one replaces the Yukawa
coupling  $y_\nu$ in equation~(\ref{eq:seesaw}) with $y_R$ in equation~(\ref{eq:lint}), that is, when the role of the SM Higgs doublet $H$ is played by $\delta^+$,
and the role of the SM lepton doublet $l_L$ is played by the SM lepton singlet $e_R$.

The baryon asymmetry of the Universe can be generated from a lepton asymmetry, as long as the latter is present before the electroweak phase transition, which occurs at a temperature $T_{EW}\simeq 150$ GeV. Above this temperature the $(B+L)$-violating 
electroweak sphalerons are in thermal equilibrium, thus converting efficiently leptons into baryons. When the SM is extended 
by sterile neutrinos, one can generate a lepton asymmetry, provided that  the set of couplings  $(M_N)_{ij}$, $(y_\nu)_{\alpha i}$ and $(y_R)_{\alpha i}$ violates the $CP$ symmetry, and (some of) these couplings are out-of-equilibrium at some epoch before $T_{EW}$.

\subsection{Leptogenesis from $N$-decays}
The traditional leptogenesis mechanism \cite{Fukugita:1986hr} assumes a Majorana mass matrix $M_N$ for two (or more) sterile neutrinos, with eigenvalues $M_i$ larger than the electroweak scale,
and it relies on the out-of-equilibrium decays $N_i \rightarrow H l_{L\alpha}$, 
at temperatures just below $M_i$. 
As $N_1$ plays the role of keV-scale dark matter  candidate, the relevant decays are those of $N_2$, with $T_{EW}<M_2<M_3$.
Since $M_2$ violates lepton number, the $CP$-asymmetry that is generated in the decays is also a lepton asymmetry, or equivalently
a $(B-L)$-asymmetry. After all the $N_2$ particles decayed, the produced ($B-L$)-asymmetry remains constant.
In this scenario, the couplings that control leptogenesis  are the same Yukawa couplings, $y_\nu$, that determine the active neutrino masses.
As a consequence,  a sufficient amount of baryon asymmetry can be generated only for $M_2 \gtrsim 10^8$ GeV \cite{Davidson:2002qv}
(barring resonance effects, that require a strong degeneracy among the sterile neutrino mass eigenstates \cite{Pilaftsis:1997jf}). 
Thus, no direct test of this scenario can be envisaged.

In the presence of a charged scalar $\delta^+$, an analog source of leptogenesis is provided by the decays 
$N_i\rightarrow \delta^+ e_{R\alpha}$, mediated by the Yukawa coupling $y_R$, as shown in \cite{Frigerio:2006gx}. Here we adapt the results of that paper to our new context, where the dark matter particle is the keV-scale sterile neutrino $N_1$. We can safely neglect the presence of $N_1$ for the computation of the lepton asymmetry because
its interaction rates (in particular those violating lepton number) are strongly out-of-equilibrium above $T_{EW}$,
due to the extreme smallness of $M_1$, $(y_R)_{\alpha 1}$ and $(y_\nu)_{\alpha 1}$. Thus, at least two extra sterile neutrinos $N_{2,3}$ heavier than $\delta^+$ are needed to generate the $CP$-asymmetry. 
Taking for simplicity $M_3\gg M_2 \gg M_\delta$, 
the $CP$-asymmetry in $N_2$-decays is given by
\begin{equation}
\epsilon_{N_2} = \frac{1}{8\pi}\frac{{\rm Im}[
\sum_{\alpha}(y_R)_{\alpha 2}(y_R)_{\alpha 3}^*]^2}{\sum_{\alpha} (y_R)_{\alpha 2}(y_R)_{\alpha 2}^*}\frac{M_2}{M_3} ~,
\end{equation}
where we assumed that the charged scalar lepton number is $L(\delta^+)=-2$, since the dominant decay mode
is $\delta^+\rightarrow e^+\overline{\nu}$, 
through the Yukawa coupling $y_L$ in equation (\ref{Ldelta}).\footnote{The asymmetry changes by
an order one factor if $y_L$ is negligibly small and $\delta^+$ undergoes slower three-body decays 
\cite{Frigerio:2006gx}.}
To reproduce the observed baryon asymmetry two basic conditions are required \cite{Frigerio:2006gx}: 
a minimal value for $\epsilon_{N_2}$ assuming no washout, 
\begin{equation}
\frac{|{\rm Im}[
\sum_{\alpha}(y_R)_{\alpha 2}(y_R)_{\alpha 3}^*]^2|}{\sum_{\alpha} (y_R)_{\alpha 2}(y_R)_{\alpha 2}^*} \gtrsim
2\cdot 10^{-6} \frac{M_3}{M_2} ~,
\end{equation}
and $N_2$-decays out-of-equilibrium at $T=M_2$ to avoid large washout from inverse decays, 
\begin{equation}
\sum_{\alpha} (y_R)_{\alpha 2}(y_R)_{\alpha 2}^* \lesssim 10^{-13} \frac{M_{2}}{1~{\rm TeV}} ~.
\end{equation}
Thus, one can realize leptogenesis for $M_2$ as small as a few TeVs, as long as 
 $N_{2,3}$ have hierarchical Yukawa couplings: $|(y_R)_{\alpha 2}|\lesssim 3\times 10^{-7}$ and 
$|(y_R)_{\alpha 3}|\gtrsim 10^{-3}\sqrt{M_3/M_2}$. 
It is worth reminding that, when $M_2$ approaches the electroweak scale, a sufficiently large CP-asymmetry implies in general 
strong $(B-L)$-washout rates, in particular those mediated by $N_3$ off-shell, unless special conditions are realized.
A neat, model-independent discussion of the lower bound on the leptogenesis scale can be found in \cite{Racker:2013lua,Racker:2014uga}. 
In the present model with $M_2\sim$ a few TeVs, the washout can be Boltzmann suppressed by raising $M_\delta$
sufficiently close to $M_2$, and taking a sufficiently large coupling $y_L$ in equation (\ref{Ldelta}) to avoid washout from the asymmetry stored in $\delta^+$.\footnote{To reduce the washout, 
one can also delay $N_2$-decays by taking $(y_R)_{\alpha 2}$ very small, but in this case one may need to produce the initial thermal
density of $N_2$ by some other interaction.}

Note that the asymmetries generated
by $N_2$-decays through the Yukawa coupling matrix $y_\nu$ become negligible in the region $M_2\ll 10^8$ GeV due to the seesaw relation. The $y_R$-entries, on the other hand, are not constrained by the light neutrino masses, 
and leptogenesis can work close to the electroweak scale as described above. This scenario is, therefore, easier to test directly,
by the observation of $\delta^+$ (and possibly $N_2$) at colliders, as discussed in section \ref{exp}.

\subsection{Leptogenesis from $N$-oscillations}
A complementary mechanism for leptogenesis \cite{Akhmedov:1998qx} relies on $N$-oscillations rather than on 
$N$-decays.
It also requires two (or more) sterile neutrinos, coupled to the SM through the Yukawa matrix $y_\nu$.
In the early Universe, the thermal population of lepton doublets $l_{L\alpha}$ produces, through small,
out-of-equilibrium couplings $(y_\nu)_{\alpha i}$, a coherent superposition $N_\alpha$ of the sterile neutrino mass eigenstates $N_i$.
These sterile neutrinos, at temperatures much larger than their masses, coherently oscillate among the different flavours $\alpha$. 
Such oscillations conserve lepton number (the $N_\alpha$ conserve their helicity), 
but violate lepton flavour numbers. If the $CP$-symmetry is also violated,
one generates
non-zero flavour asymmetries between the opposite helicities of the  $N_\alpha$.
The asymmetry in the flavour $\alpha$ is transferred efficiently to $l_{L\alpha}$, 
as long as the coupling $(y_{\nu})_{\alpha i}$  goes into equilibrium for some $i$.
Since the total lepton asymmetry, that is, the sum over $\alpha$ of the flavour asymmetries vanishes,
one needs that some but not all flavours go into equilibrium before $T_{EW}$, 
so that a net lepton asymmetry remains stored in the sterile neutrino sector, 
and an opposite one is available in the SM sector to be transferred to the baryons by electroweak sphalerons.
The Yukawa (out-of-)equilibrium condition at $T_{EW}$ reads
\begin{equation}
|(y_\nu)_{\alpha i}| \gtrsim 10^{-7} {\rm ~for~some~} \alpha {\rm~and~} i~,~
|(y_\nu)_{\beta i}| \lesssim 10^{-7} {\rm ~for~some~} \beta\ne\alpha,~\forall i ~.
\label{flavourWashout}\end{equation}
A number of additional constraints, on the size of $y_\nu$-entries and on the values of $M_i$, must be satisfied  for this leptogenesis mechanism to work. To facilitate the comparison between the scenarios with and without the charged scalar, we will describe them in some detail.  Due to the dark matter constraint,  $N_1$  plays no role in  
leptogenesis and the asymmetry must be generated by $N_2$ and $N_3$ only.\footnote{
Note that the seesaw lagrangian can violate $CP$ with only two sterile neutrinos, as it contains three
physical phases, one combination of them being relevant in oscillations}.

 Successful leptogenesis implies a few upper bounds on the $y_\nu$-entries, beside the flavour-dependent one in equation~(\ref{flavourWashout}).
First of all, note that the Majorana masses $M_i$ violate lepton number.
Therefore, they play the role of washout in this scenario, as they transform opposite helicities of the $N_\alpha$ into one another. 
The condition to keep the lepton number violation rate out-of-equilibrium down to $T_{EW}$ is approximately
\begin{equation}
|(y_\nu)_{\alpha i}| \lesssim 10^{-5} ~\frac{\rm GeV}{M_i},~\forall \alpha~{\rm and}~\forall i~, 
\label{leptonWashout}\end{equation}
which combined with equation (\ref{flavourWashout}) implies sterile neutrino masses 
below the electroweak scale, $M_i\lesssim 100$ GeV.\footnote{We remark that most numerical studies 
in the literature assume that the lepton number violating rates are negligible; in view of the size of the Yukawa couplings relevant for leptogenesis,
this assumptions seems to be justified only for $M_i \lesssim$ GeV. For larger $M_i$, one should include the lepton number violating rates
in the Boltzmann equations. In this regime, it remains conceivable that
the $CP$-asymmetries are large enough to compensate the relatively strong washout. This possibility was recently entertained 
to make this scenario work even for $M_i$ larger than $T_{EW}$ \cite{Garbrecht:2014bfa}.}
In addition, for the case of two sterile neutrinos $N_{2,3}$,
analytic and numerical studies \cite{Asaka:2005pn,Canetti:2010aw,Shuve:2014zua} show that a sufficient baryon asymmetry requires 
a strong degeneracy between $M_2$ and $M_3$, with $\Delta M/M\lesssim 10^{-5}$.
This comes from the interplay of a number of subtle effects: the oscillation time increases as $\Delta M$ decreases, and this allows for larger 
asymmetries because the Yukawa interaction rates are closer to equilibrium at later times; note also that larger Yukawa 
couplings (for some but not all the flavours) enhance the flavour asymmetries, but also tend to spoil the coherence of the sterile neutrinos, as
active-sterile transitions may become faster than the oscillation time. We extrapolate the resulting constraint from
figure 7 of \cite{Canetti:2012kh}, that can be written as 
\begin{equation}
|(y_\nu)_{\alpha i}| \lesssim 2\cdot 10^{-6}\left(\frac{M_i}{\rm GeV}\right)^{1/2}~\forall \alpha~{\rm and}~\forall i~.
\label{extrap}\end{equation}

The masses $M_{2,3}$ are below collider energies, but it is difficult to produce them directly since they have small 
couplings.  
Still, the mixing with active neutrinos can be sufficiently
large to have an observable effect in various neutrino experiments \cite{Gorbunov:2007ak}. 
The resulting upper bound on active-sterile mixing is also shown in figure 7 of \cite{Canetti:2012kh}.
 An experimental proposal to improve significantly the present bound can be  found in \cite{Bonivento:2013jag}.
In less minimal models, e.g. with three GeV scale sterile neutrinos, there are good detection perspectives in meson decay experiments \cite{Canetti:2014dka}.

On the other hand, several observables other than the baryon asymmetry put lower bounds on the $y_\nu$-entries, 
that add to the one in equation~(\ref{flavourWashout}).
The lower bound on active neutrino masses, $|(m_\nu)_{\alpha\beta}|\gtrsim 0.025$ eV for some $\alpha$ and $\beta$, implies a lower bound on the neutrino Yukawa couplings
through the seesaw formula,
\begin{equation}
|(y_\nu)_{\alpha i}| \gtrsim 2\cdot 10^{-8}\left(\frac{M_i}{\rm GeV}\right)^{1/2} {\rm ~for~some~} \alpha {\rm~and~} i~.
\label{seesawBound}\end{equation}
Indeed, since $N_1$ gives a negligible contribution to $m_\nu$, one active neutrino is approximately massless,
therefore  $|(m_\nu)_{\alpha\beta}|\lesssim 0.05$ eV for all $\alpha$ and $\beta$. Then,
the right-hand side of equation (\ref{seesawBound}) provides the natural value of the Yukawa couplings  for $i=2,3$: the largest
$(y_\nu)_{\alpha i}$ are, the strongest is the cancellation needed among the contributions of $N_2$ and $N_3$ to $m_\nu$.
These GeV-scale sterile neutrinos should decay (e.g. in $3\nu$'s or $\nu e^+e^-$) before $T\simeq 1$ MeV, not to spoil 
nucleosynthesis (see \cite{Ruchayskiy:2012si} for a detailed analysis). We roughly estimate this constraint as 
\begin{equation}
|(y_\nu)_{\alpha i}| \gtrsim 3\cdot 10^{-8}\left(\frac{\rm GeV}{M_i}\right)^{3/2},{\rm ~for~some~} \alpha, ~\forall i~.
\label{BBNbound}\end{equation}
Comparing with equation (\ref{extrap}), this implies $M_i\gtrsim 0.1$ GeV. All the constraints above confirm that $N_1$ plays no role in leptogenesis.

Let us show  that the same mechanism of leptogenesis through $N$-oscillations is operative in the presence of the
charge scalar $\delta^+$, by replacing the role of $y_\nu$ with $y_R$. Analogously to the previous case, 
the couplings $(y_R)_{\alpha i}$ must be small to remain out-of-equilibrium 
while they slowly produce coherent sterile neutrino states.   
The lepton flavour asymmetries generated by $N$-oscillations are (partially) transferred to the SM lepton singlets 
$e_{R\alpha}$.
The latter are in equilibrium with $l_{L\alpha}$ through the charged lepton Yukawa couplings (at least for 
$\alpha=\mu,\tau$), thus electroweak sphalerons transfer
the asymmetries to baryons as usual. 

One needs that the $y_R$-entries satisfy the same inequalities as the $y_\nu$-entries in equations  (\ref{flavourWashout})
and (\ref{leptonWashout}),
so that some flavour asymmetries are transferred to baryons, while the others remain stored in the sterile sector,
and the washout from lepton number violating scattering is small.
The only differences amount to (i) order one factors to account for the singlet (doublet) nature of $e_R$ ($l_L$) in scattering rates;
(ii) the  range of temperatures where $M_\delta$ ($M_H$) can be neglected: note that the asymmetry is generated at some scale well above 
$T_{EW}$ through scattering processes mediated by $\delta^+$ ($H$).

We also expect that, when both $y_\nu$ and $y_R$ take values relevant for leptogenesis, there is more freedom to generate
large asymmetries, e.g. because of the presence of extra $CP$-violating phases, and the requirement of a strong degeneracy
between $M_2$ and $M_3$ could be relaxed. 
As in the case of $y_\nu$, one cannot raise  too much the value of $y_R$-entries, as the coherence of sterile neutrino oscillations requires the scattering rate with the SM leptons to be out-of-equilibrium at the time of oscillations; this should translate in a bound
 similar to the one in equation (\ref{extrap}).  A numerical study 
is needed to establish more precisely the allowed parameter space, and the differences with respect to the $\nu$MSM one.
The effect of the two sets of Yukawa couplings could be dramatic, as is the effect of a third GeV scale sterile neutrino
\cite{Drewes:2012ma,Canetti:2014dka}, or of a second Higgs doublet \cite{Shuve:2014zua}.

There is no constraint on $y_R$ coming from the seesaw relation, so one can take couplings smaller than in 
equation (\ref{seesawBound}), slightly enlarging the region of parameters of leptogenesis. Perhaps more importantly, one
can take $(y_R)_{\alpha i}$ significantly larger than the right-hand side of equation (\ref{seesawBound}), with no need of
fine-tuning to keep $m_\nu$ small. Note that the coupling $y_R$ does not induce any mixing with active neutrinos.
Indeed, direct searches of active-sterile mixing are presently sensitive to $y_\nu$-entries 
much larger than in equation (\ref{seesawBound}): if the seesaw parameters take their natural values, 
no direct signal of active-sterile mixing is expected. 

Finally, the three-body decay rate of $N_i$  through the coupling $y_R$ is proportional to 
$|(y_R)_{\alpha i} (y_L)_{\beta\gamma}|^2/M_\delta^4$, to be compared with $|(y_\nu)_{\alpha i}|^2/(M_i^2 v^2)$
for a decay through the mixing with active neutrinos.
Since the latter is typically much faster, the nucleosynthesis bound applies to $y_\nu$  only, in the form of equation (\ref{BBNbound}).
Then, leptogenesis through $y_R$ could work 
 even for $M_i < 0.1$ GeV, but  in this region nucleosynthesis demands large values for the $y_\nu$-entries,
that require strong cancellations in the seesaw.

\section{Experimental signatures for   $N_1$ and $\delta^+$ \label{exp}} 
Two important differences between our model and the  $\nu$MSM are the additional region of the parameter space where the dark matter constraint can be satisfied and the presence of the extra charged scalar $\delta^+$. They both give rise to new experimental signatures that may allow to distinguish one model from the other. 

\subsection{Dark matter indirect detection}

A multi-keV dark matter neutrino is inherently unstable and decays into three light neutrinos at tree-level ($N_1\to 3\nu_\alpha$) 
and radiatively into a light neutrino and a photon ($N_1\to \nu_\alpha\gamma$). This radiative decay produces an X-ray line at  $E_\gamma  \simeq  M_{1}/2$ that can be searched for and used to constrain the model or to help determine its parameter space \cite{Abazajian:2001vt}. The presence of the additional scalar $\delta^+$ does not affect the decay modes of the dark matter neutrino, which are still determined by its mass and its mixing with the active neutrinos, just as in the $\nu$MSM, but it modifies the regions that are consistent with the dark matter constraint, allowing, in particular, for smaller mixing angles. This fact has important implications, as we show next.
 
Recently, the detection of an unidentified spectral line at about $3.5$ keV has been reported  
from two independent data sets \cite{Bulbul:2014sua,Boyarsky:2014jta}. Arguments in favour or against the dark matter decay interpretation of the signal can be found
in \cite{Boyarsky:2014ska,Jeltema:2014qfa,Malyshev:2014xqa,Boyarsky:2014paa}. 
If confirmed, that signal would provide compelling evidence for keV-scale dark matter and, in particular, for dark matter in the form of sterile neutrinos. Within that framework, the signal can be explained if  $M_1\simeq  7$ keV and $\sin^22\theta_1\simeq 5\times10^{-11}$.
In the minimal scenario for sterile neutrino dark matter, where they are produced non-resonantly, such parameters lead to a relic density way below the range 
determined by  cosmological observations. Thus, an additional source of sterile neutrinos is required. A simple possibility for that new source are the  decays of the $\delta^+$ particle, as explained in the previous section. 
These decays allow to decouple the  dark matter production (determined by the $y_R$ couplings) from the dark matter decay (determined by $y_\nu$ via mixing angles), opening
 new viable regions consistent with all bounds. 
Specifically, the freeze-in production of sterile neutrinos that we have examined in this paper enables to explain the tentative $3.5$ keV line and to simultaneously account for the observed relic density, even in the absence of a large, primordial lepton asymmetry. If that line signal turns out to be spurious,  
X-ray observations will continue to be the  main way in which the dark matter sector of this model can be tested in the foreseeable future.

\subsection{Charged scalar searches at colliders}

The singly-charged isosinglet scalar $\delta^+$ could be directly produced at colliders, if its mass were within their energy range. The dominant production channel is the Drell-Yan process $\psi\overline{\psi}\rightarrow
\gamma/Z\rightarrow \delta^+\delta^-$, with $\psi=e$ at LEP and  $\psi=q$ at Tevatron and  LHC, with a partonic cross-section given e.g. in \cite{Muhlleitner:2003me}. 
The $\delta^+$ decays into one anti-lepton and one anti-neutrino, either through the Yukawa coupling $(y_L)_{\alpha\beta}$, or 
$(y_R)_{\alpha i}$ when $M_i<M_\delta$.\footnote{
In principle these couplings can all be very small, possibly leading to dominant three-body decays into $l^+l^-W^+$ \cite{Frigerio:2006gx}. 
In this case $\delta^+$ can be sufficiently long-lived to  appear as a curved charged track across the whole detector.}

The standard experimental searches for charged scalars usually assume production and decay modes different from those above. 
For a singly-charged scalar $H^+$, the decay channel into lepton plus missing energy has been analyzed, but only for an isodoublet produced from a top-bottom vertex, as in type II two-Higgs doublet models, allowing for an effective background reduction \cite{Chatrchyan:2012vca,ATLAS-CONF-2013-090}.
Indeed, the signal over background ratio is significantly smaller in the case of $\delta^+$;
an analysis of the cuts required to maximize the signal has been presented  in section 5.2 of \cite{DelNobile:2009st}. 

Presently, it appears that the best way to constrain $M_\delta$ is to use the experimental searches for supersymmetric particles. Since $\delta^+$ and a right-handed slepton have the same gauge quantum numbers,  when the latter is 
directly pair-produced and decays into a lepton and a light neutralino, it behaves very much as the former, and the same bounds apply to both particles.  
Recently, that setup was studied in the context of simplified supersymmetric models, allowing to extract a 95\% C.L. bound
$M_\delta \gtrsim 250$ GeV at ATLAS \cite{Aad:2014vma} and $M_\delta \gtrsim 190$ GeV at CMS \cite{Khachatryan:2014qwa}.

\section{Flavour structure  of the model\label{flavour}}

We accomplished our goal of demonstrating that this model can account for dark matter, neutrino masses, and baryogenesis. That is, we have determined, in the previous sections, that there  exist regions in the parameter space of the model where the masses of the new particles and their  Yukawa couplings are such that  these three issues are simultaneously explained. One may wonder, nonetheless, whether those values of the masses and Yukawa couplings 
have a generic flavour structure, or if they require a special tuning, indicating that some flavour symmetries are operative. Previous works along this line include \cite{Merle:2011yv,Barry:2011fp,Shaposhnikov:2006nn,Lindner:2010wr,Merle:2013gea}, which focused on the generation of the keV scale in neutrino models, and on the flavour structure of the $\nu$MSM. In this section, we do not aim to construct complete flavour models, but rather to recap the order of magnitude of the parameters needed phenomenologically, and to suggest some rationale to explain them.

In view of the hierarchical values of masses and couplings that are required for dark matter and leptogenesis, it is useful to describe the required flavour structure in terms
of a $U(1)_F$ family symmetry, with different charges assigned to the various fermions. In this framework, the coefficient of each fermion bilinear $\psi_{Li}\psi_{Lj}$ is suppressed by a power $q_i+q_j$ of a small parameter $\epsilon$ (we take the fermion $U(1)_F$-charges to be positive). 
One may assume some underlying flavour dynamics, that generates 
$\epsilon=\langle\phi\rangle/\Lambda_F\ll 1$,  that is, the vev of a spurion field 
with $q_\phi = -1$, over the cutoff of the flavour theory.

We need $M_1\ll M_2\sim M_3$. Note that in the $\nu$MSM a strong degeneracy of $M_2$ and $M_3$ is required, while in our scenario this constraint is relaxed.
The straightforward way to realize this pattern is to take $q_{N1} > q_{N2}=q_{N3}$, so that $M_1/M_{2}\sim \epsilon^{2n}$, with
$n\equiv q_{N1}-q_{N2}$. For $N$-oscillation ($N$-decay) leptogenesis, one needs $\epsilon^{2n} \sim 10$ keV/GeV $=10^{-5}$ 
($\epsilon^{2n}\lesssim 10$ keV/TeV $= 10^{-8}$). This charge assignment implies automatically a hierarchy $\epsilon^n : 1 : 1$ among the three columns of 
the matrix $y_R$ (as well as of $y_\nu$), in other words, $(y_R)^2_{\alpha i} /(y_R)^2_{\alpha j} \sim M_i/M_j$.
Since this relation is not satisfied in certain regions of parameters relevant for dark matter, leptogenesis and neutrino masses, an additional flavour symmetry must be operative. The Yukawa couplings can be further suppressed
by introducing parity symmetries $Z_2^{(i)}:~N_i\rightarrow -N_i$, that allow for $M_i$ but forbid $(y_\nu)_{\alpha i}$ and $(y_R)_{\alpha i}$;
these couplings must then be proportional to a small symmetry-breaking parameter $\epsilon_i$.
To suppress $(y_\nu)_{\alpha i}$ and not $(y_R)_{\alpha i}$ (or vice versa), one may argue that under the same parity the combination 
$e_{R\alpha}\delta^+$ (or $l_{L\alpha} H$) is also odd.\footnote{
The other way around, one could also enhance the hierarchy among the $M_i$ relatively to the hierarchy among the columns of $y_\nu$ and $y_R$, 
by introducing a (family-dependent) lepton number $U(1)_L$, that is conserved in the Yukawa couplings and broken by two units in the Majorana mass term.
We will not need such a symmetry in the following.}

Coming to the $U(1)_F$
charges of the three families of lepton doublets and singlets,  $q_{L\alpha}$ and $q_{R\alpha}$, first of all they
determine the hierarchy of the charged lepton masses, 
$m_e : m_\mu : m_\tau\sim \epsilon^{q_{Le}+q_{Re}} :\epsilon^{q_{L\mu}+q_{R\mu}}: \epsilon^{q_{L\tau}+q_{R\tau}}$. 
In turn, the hierarchy among the rows of $y_R$ ($y_\nu$) is determined by the charges $q_{R\alpha}$ ($q_{L\alpha}$). Note that the charges of 
$l_{L\alpha}$ and $e_{R\alpha}$ are important for charged lepton and active neutrino masses, as well as for 
leptogenesis from $N$-oscillations; however they are not very relevant for dark matter production nor for leptogenesis from $N$-decays.

Let us confront these simple flavour symmetries with the values of the parameters needed for dark matter, 
neutrino masses and  leptogenesis. The freeze-in of the desired amount of $N_1$ 
from  $\delta^+$ decays requires $\sum_\alpha |(y_R)_{\alpha1}|^2 \simeq 2.5\times 10^{-24} M_\delta/M_1$
(see equation (\ref{relic})). The $N_1$-production from active-sterile oscillations is negligible for
$|(y_\nu)_{\alpha 1}| \lesssim 10^{-13} (M_1/{\rm keV})$ (barring a large primordial lepton asymmetry). To generate large enough active neutrino masses, equation (\ref{seesawBound}) must be satisfied. Clearly $N_1$ does not contribute
significantly to $m_\nu$ because of the dark matter constraint, therefore $N_2$ and $N_3$ are responsible to generate the atmospheric and solar mass scales.
Since $\sqrt{\Delta m^2_{sol}/\Delta m^2_{atm}}\simeq 0.2$, equation (\ref{seesawBound}) must hold for both $i=2$ and $3$, up to a factor of a few.
Coming to leptogenesis, the two mechanisms discussed in section \ref{barlep} correspond to two 
very different regions of parameters that we discuss in turn.

In the case of leptogenesis from  $N$-decays through the $y_R$-interaction, one requires $M_{2,3}>$ TeV, as well as 
$|(y_R)_{\alpha 3}| > 10^{-3}\sqrt{M_3/M_2}$ for
at least one $\alpha$, and $|(y_R)_{\alpha 2}| < 10^{-4} \sqrt{M_2/(10^8{\rm~GeV})}$ for all $\alpha$.
For definiteness, let us take $M_1\sim$ keV, $M_\delta \sim$ TeV and $M_{2,3} \sim 100 $ TeV
 (leptogenesis through $y_\nu$ is irrelevant at such low scales, barring resonances).
Taking $\epsilon=0.1$ and the $U(1)_F$ charges
$q_{N1}=7$, $q_{N2}=q_{N3}=2$, $q_{R\tau}=0\ge q_{R\mu},q_{Re}$,  $q_{L\tau}=2 \ge q_{L\mu},q_{Le}$, 
we can reproduce the correct size of the   $M_N$, $y_R$ and $y_\nu$ entries, except for the second column of $y_R$,
that must be further suppressed by  $\epsilon_{R2}\lesssim 10^{-3}$, and the first column of $y_\nu$,
to be further suppressed by $\epsilon_{\nu1}\lesssim 10^{-4}$.
As discussed above, $\epsilon_{R2}$ ($\epsilon_{\nu1}$) can be associated to the breaking of $N_2$-parity ($N_1$-parity). Note that here $M_{2,3}\sim \epsilon^4 \Lambda_L$, with a lepton number violation scale $\Lambda_L \sim 10^9$ GeV.

In the case of leptogenesis from $N$-oscillations coupled to the SM through $y_R$, one requires for $i=2,3$, $M_{i}<100$ GeV,
$|(y_R)_{\alpha i}|M_i\lesssim 10^{-5}$ GeV, as well as $|(y_R)_{\alpha i}| > 10^{-7}$ for some flavour
 $\alpha$ and $< 10^{-7}$ for some different flavour  $\beta$.
For definiteness, let us take $M_1\sim 10 $ keV,  $M_{2,3} \sim 10$ GeV and $M_\delta \sim$ TeV.
Then, one can reproduce the correct size of the $M_N$ and $y_R$ entries taking $\epsilon=0.1$ and $U(1)_F$ charges
$q_{N1}=8$,  $q_{N2}=q_{N3}= 5$, $q_{R\tau}=1$, $q_{R\mu}=2$ and $q_{Re}=3$,  and charged lepton masses 
require $q_{L\alpha} = 1 \ge q_{L\mu},q_{Le}$. 
Coming to the structure of $y_\nu$, one needs an extra source of suppression for $(y_\nu)_{\alpha 1}$, by a factor $\epsilon_{\nu 1} \lesssim 10^{-3}$,
to avoid $N_1$-overproduction from active-sterile mixing;
as before, $\epsilon_{\nu 1}$ can be understood as the breaking parameter of $N_1$-parity.
We note that, for the chosen values of parameters,  both $y_R$ and $y_\nu$ couplings to $N_{2,3}$ are relevant for leptogenesis.
Here lepton number is broken at the scale $\Lambda_L \sim M_{2,3}/\epsilon^{10}\sim10^{11}$ GeV.

In summary, the regions of parameters where our scenario is effective require a well-defined hierarchy of masses and couplings,
that point to specific but relatively natural flavour structures, with no need of extreme tuning of parameters. 

\section{Conclusions}
\label{sec:con}

We have shown that, when the SM is extended with three sterile neutrinos $N_i$ and a charged scalar $\delta^+$, baryogenesis via leptogenesis and the production of dark matter  can both be mediated by  $\delta^+$.
 
The dark matter candidate in this scenario is the lightest  sterile neutrino $N_1$, with a mass  $M_1$ larger than about a few keVs. 
Such  sterile neutrinos are produced by the decays of $\delta^+$ while it is in thermal equilibrium in the early Universe. We showed that this novel mechanism, which is an example of freeze-in dark matter production, can account for the observed dark matter density. Moreover, since the resulting relic density does not depend on the active-sterile mixing angle
$\theta_1$, the constraints from X-ray data  are alleviated. 
The recent hint of a signal at $3.5$ keV can be easily explained by $N_1$-decays, for $\theta_1\sim 3\times 10^{-6}$. 
Alternatively, for a vanishingly small $\theta_1$, the dark matter mass $M_1$ can be raised to the MeV scale, or beyond.

Leptogenesis can be realized  through the coupling $y_R$ between $\delta^+$ and $N_{2,3}$ in two different ways, either via  $N$-decays
or via $N$-oscillations. Both mechanisms are well-known in the context of the SM plus sterile neutrinos only, as they can both proceed through the neutrino Yukawa coupling $y_\nu$. 
The latter is constrained by light neutrino masses, restricting significantly the allowed parameter space and the testability of leptogenesis.
In contrast, in our scenario $y_R$ is not involved in the neutrino mass generation.
As a consequence, the $N$-decay mechanism is successful for $M_{2,3}$ as small as a few TeVs, with no need of a resonant enhancement.
The parameter window for $N$-oscillation leptogenesis is similar when one employs the coupling matrix $y_R$ instead of $y_\nu$:
one needs roughly  $M_{2,3}\sim$  GeV and $y_R$-entries smaller than $10^{-6}$. However, $y_R$ introduces additional
sources of $CP$-violation, and moreover it is not constrained neither by light neutrino masses nor by nucleosynthesis. Thus one can largely relieve
the fine-tuning of parameters necessary in the case with $y_\nu$ only.

Neutrino masses are generated by a low scale seesaw mechanism involving the two heavier singlets, $N_{2,3}$,  the $N_1$-contribution being negligible. 
In this way one can accommodate the current data on neutrino masses and mixing angles,  with the prediction of an almost massless lightest neutrino.
At colliders such as the LHC, this scenario can be probed and constrained via the direct production of  $\delta^+$. 
This minimal extension of the SM thus provides a simple and testable scenario to explain neutrino masses, the dark matter, and the baryon asymmetry of the Universe.

\section*{Acknowledgments}
MF thanks F.~Bezrukov, M.~Drewes, T.~Hambye, K.~Jedamzik and J.~Racker for useful discussions.
MF is partially supported by the FP7 European ITN project ``Invisibles" (PITN-
GA-2011-289442-INVISIBLES), and by the OCEVU Labex (ANR-11-LABX-0060) 
funded by the "Investissements d'Avenir" French government program managed by the ANR.
C.Y. is partially supported by the ``Helmholtz Alliance for Astroparticle Physics HAP''
 funded by the Initiative and Networking Fund of the Helmholtz Association.

\bibliographystyle{hunsrt}
\bibliography{darkmatter}

\begin{thebibliography}{10}

\bibitem{Dodelson:1993je}
Scott Dodelson and Lawrence~M. Widrow.
\newblock {Sterile-neutrinos as dark matter}.
\newblock {\em Phys.Rev.Lett.}, 72:17--20, 1994, hep-ph/9303287.

\bibitem{Fukugita:1986hr}
M.~Fukugita and T.~Yanagida.
\newblock {Baryogenesis Without Grand Unification}.
\newblock {\em Phys.Lett.}, B174:45, 1986.

\bibitem{Akhmedov:1998qx}
Evgeny~K. Akhmedov, V.A. Rubakov, and A.~Yu. Smirnov.
\newblock {Baryogenesis via neutrino oscillations}.
\newblock {\em Phys.Rev.Lett.}, 81:1359--1362, 1998, hep-ph/9803255.

\bibitem{Asaka:2005pn}
Takehiko Asaka and Mikhail Shaposhnikov.
\newblock {The nuMSM, dark matter and baryon asymmetry of the universe}.
\newblock {\em Phys.Lett.}, B620:17--26, 2005, hep-ph/0505013.

\bibitem{Asaka:2005an}
Takehiko Asaka, Steve Blanchet, and Mikhail Shaposhnikov.
\newblock {The nuMSM, dark matter and neutrino masses}.
\newblock {\em Phys.Lett.}, B631:151--156, 2005, hep-ph/0503065.

\bibitem{Canetti:2012kh}
Laurent Canetti, Marco Drewes, Tibor Frossard, and Mikhail Shaposhnikov.
\newblock {Dark Matter, Baryogenesis and Neutrino Oscillations from Right
  Handed Neutrinos}.
\newblock {\em Phys.Rev.}, D87:093006, 2013, 1208.4607.

\bibitem{Abazajian:2006yn}
Kevork Abazajian and Savvas~M. Koushiappas.
\newblock {Constraints on Sterile Neutrino Dark Matter}.
\newblock {\em Phys.Rev.}, D74:023527, 2006, astro-ph/0605271.

\bibitem{Boyarsky:2007ay}
Alexey Boyarsky, Dmytro Iakubovskyi, Oleg Ruchayskiy, and Vladimir Savchenko.
\newblock {Constraints on decaying Dark Matter from XMM-Newton observations of
  M31}.
\newblock {\em Mon.Not.Roy.Astron.Soc.}, 387:1361, 2008, 0709.2301.

\bibitem{Horiuchi:2013noa}
Shunsaku Horiuchi, Philip~J. Humphrey, Jose Onorbe, Kevork~N. Abazajian, Manoj
  Kaplinghat, et~al.
\newblock {Sterile neutrino dark matter bounds from galaxies of the Local
  Group}.
\newblock {\em Phys.Rev.}, D89:025017, 2014, 1311.0282.

\bibitem{Boyarsky:2008xj}
Alexey Boyarsky, Julien Lesgourgues, Oleg Ruchayskiy, and Matteo Viel.
\newblock {Lyman-alpha constraints on warm and on warm-plus-cold dark matter
  models}.
\newblock {\em JCAP}, 0905:012, 2009, 0812.0010.

\bibitem{Viel:2006kd}
Matteo Viel, Julien Lesgourgues, Martin~G. Haehnelt, Sabino Matarrese, and
  Antonio Riotto.
\newblock {Can sterile neutrinos be ruled out as warm dark matter candidates?}
\newblock {\em Phys.Rev.Lett.}, 97:071301, 2006, astro-ph/0605706.

\bibitem{Viel:2013fqw}
M.~Viel, G.D. Becker, J.S. Bolton, and M.G. Haehnelt.
\newblock {Warm Dark Matter as a solution to the small scale crisis: new
  constraints from high redshift Lyman-alpha forest data}.
\newblock {\em Physical Review D, vol. 88, Issue 4, id.}, 043502, 2013,
  1306.2314.

\bibitem{Shi:1998km}
Xiang-Dong Shi and George~M. Fuller.
\newblock {A New dark matter candidate: Nonthermal sterile neutrinos}.
\newblock {\em Phys.Rev.Lett.}, 82:2832--2835, 1999, astro-ph/9810076.

\bibitem{Laine:2008pg}
M.~Laine and M.~Shaposhnikov.
\newblock {Sterile neutrino dark matter as a consequence of nuMSM-induced
  lepton asymmetry}.
\newblock {\em JCAP}, 0806:031, 2008, 0804.4543.

\bibitem{Shaposhnikov:2008pf}
Mikhail Shaposhnikov.
\newblock {The nuMSM, leptonic asymmetries, and properties of singlet
  fermions}.
\newblock {\em JHEP}, 0808:008, 2008, 0804.4542.

\bibitem{Shuve:2014zua}
Brian Shuve and Itay Yavin.
\newblock {Baryogenesis through Neutrino Oscillations: A Unified Perspective}.
\newblock {\em Phys.Rev.}, D89:075014, 2014, 1401.2459.

\bibitem{Shaposhnikov:2006xi}
Mikhail Shaposhnikov and Igor Tkachev.
\newblock {The nuMSM, inflation, and dark matter}.
\newblock {\em Phys.Lett.}, B639:414--417, 2006, hep-ph/0604236.

\bibitem{Bezrukov:2009yw}
F.~Bezrukov and D.~Gorbunov.
\newblock {Light inflaton Hunter's Guide}.
\newblock {\em JHEP}, 1005:010, 2010, 0912.0390.

\bibitem{Bezrukov:2014nza}
F.~Bezrukov and D.~Gorbunov.
\newblock {Relic Gravity Waves and 7 keV Dark Matter from a GeV scale
  inflaton}.
\newblock 2014, 1403.4638.

\bibitem{Bezrukov:2009th}
F.~Bezrukov, H.~Hettmansperger, and M.~Lindner.
\newblock {keV sterile neutrino Dark Matter in gauge extensions of the Standard
  Model}.
\newblock {\em Phys.Rev.}, D81:085032, 2010, 0912.4415.

\bibitem{Kusenko:2006rh}
Alexander Kusenko.
\newblock {Sterile neutrinos, dark matter, and the pulsar velocities in models
  with a Higgs singlet}.
\newblock {\em Phys.Rev.Lett.}, 97:241301, 2006, hep-ph/0609081.

\bibitem{Merle:2013wta}
Alexander Merle, Viviana Niro, and Daniel Schmidt.
\newblock {New Production Mechanism for keV Sterile Neutrino Dark Matter by
  Decays of Frozen-In Scalars}.
\newblock {\em JCAP}, 1403:028, 2014, 1306.3996.

\bibitem{Frigerio:2006gx}
Michele Frigerio, Thomas Hambye, and Ernest Ma.
\newblock {Right-handed sector leptogenesis}.
\newblock {\em JCAP}, 0609:009, 2006, hep-ph/0603123.

\bibitem{Asaka:2006nq}
Takehiko Asaka, Mikko Laine, and Mikhail Shaposhnikov.
\newblock {Lightest sterile neutrino abundance within the nuMSM}.
\newblock {\em JHEP}, 0701:091, 2007, hep-ph/0612182.

\bibitem{Roy:2010xq}
Ananda Roy and Mikhail Shaposhnikov.
\newblock {Resonant production of the sterile neutrino dark matter and
  fine-tunings in the [nu]MSM}.
\newblock {\em Phys.Rev.}, D82:056014, 2010, 1006.4008.

\bibitem{Hall:2009bx}
Lawrence~J. Hall, Karsten Jedamzik, John March-Russell, and Stephen~M. West.
\newblock {Freeze-In Production of FIMP Dark Matter}.
\newblock {\em JHEP}, 1003:080, 2010, 0911.1120.

\bibitem{Klasen:2013ypa}
Michael Klasen and Carlos~E. Yaguna.
\newblock {Warm and cold fermionic dark matter via freeze-in}.
\newblock {\em JCAP}, 1311:039, 2013, 1309.2777.

\bibitem{Molinaro:2014lfa}
Emiliano Molinaro, Carlos~E. Yaguna, and Oscar Zapata.
\newblock {FIMP realization of the scotogenic model}.
\newblock {\em JCAP}, 1407:015, 2014, 1405.1259.

\bibitem{Petraki:2007gq}
Kalliopi Petraki and Alexander Kusenko.
\newblock {Dark-matter sterile neutrinos in models with a gauge singlet in the
  Higgs sector}.
\newblock {\em Phys.Rev.}, D77:065014, 2008, 0711.4646.

\bibitem{Kusenko:2009up}
Alexander Kusenko.
\newblock {Sterile neutrinos: The Dark side of the light fermions}.
\newblock {\em Phys.Rept.}, 481:1--28, 2009, 0906.2968.

\bibitem{Petraki:2008ef}
Kalliopi Petraki.
\newblock {Small-scale structure formation properties of chilled sterile
  neutrinos as dark matter}.
\newblock {\em Phys.Rev.}, D77:105004, 2008, 0801.3470.

\bibitem{Kusenko:2013saa}
Alexander Kusenko and Leslie~J. Rosenberg.
\newblock {Working Group Report: Non-WIMP Dark Matter}.
\newblock 2013, 1310.8642.

\bibitem{Davidson:2002qv}
Sacha Davidson and Alejandro Ibarra.
\newblock {A Lower bound on the right-handed neutrino mass from leptogenesis}.
\newblock {\em Phys.Lett.}, B535:25--32, 2002, hep-ph/0202239.

\bibitem{Pilaftsis:1997jf}
Apostolos Pilaftsis.
\newblock {CP violation and baryogenesis due to heavy Majorana neutrinos}.
\newblock {\em Phys.Rev.}, D56:5431--5451, 1997, hep-ph/9707235.

\bibitem{Racker:2013lua}
J.~Racker.
\newblock {Mass bounds for baryogenesis from particle decays and the inert
  doublet model}.
\newblock {\em JCAP}, 1403:025, 2014, 1308.1840.

\bibitem{Racker:2014uga}
J.~Racker and N.~Rius.
\newblock {Helicitogenesis: WIMPy baryogenesis with sterile neutrinos and other
  realizations}.
\newblock 2014, 1406.6105.

\bibitem{Garbrecht:2014bfa}
Bjorn Garbrecht.
\newblock {More Viable Parameter Space for Leptogenesis}.
\newblock 2014, 1401.3278.

\bibitem{Canetti:2010aw}
Laurent Canetti and Mikhail Shaposhnikov.
\newblock {Baryon Asymmetry of the Universe in the NuMSM}.
\newblock {\em JCAP}, 1009:001, 2010, 1006.0133.

\bibitem{Gorbunov:2007ak}
Dmitry Gorbunov and Mikhail Shaposhnikov.
\newblock {How to find neutral leptons of the $\nu$MSM?}
\newblock {\em JHEP}, 0710:015, 2007, 0705.1729.

\bibitem{Bonivento:2013jag}
W.~Bonivento, A.~Boyarsky, H.~Dijkstra, U.~Egede, M.~Ferro-Luzzi, et~al.
\newblock {Proposal to Search for Heavy Neutral Leptons at the SPS}.
\newblock 2013, 1310.1762.

\bibitem{Canetti:2014dka}
Laurent Canetti, Marco Drewes, and Björn Garbrecht.
\newblock {Lab-to-Genesis}.
\newblock 2014, 1404.7114.

\bibitem{Ruchayskiy:2012si}
Oleg Ruchayskiy and Artem Ivashko.
\newblock {Restrictions on the lifetime of sterile neutrinos from primordial
  nucleosynthesis}.
\newblock {\em JCAP}, 1210:014, 2012, 1202.2841.

\bibitem{Drewes:2012ma}
Marco Drewes and Björn Garbrecht.
\newblock {Leptogenesis from a GeV Seesaw without Mass Degeneracy}.
\newblock {\em JHEP}, 1303:096, 2013, 1206.5537.

\bibitem{Abazajian:2001vt}
Kevork Abazajian, George~M. Fuller, and Wallace~H. Tucker.
\newblock {Direct detection of warm dark matter in the X-ray}.
\newblock {\em Astrophys.J.}, 562:593--604, 2001, astro-ph/0106002.

\bibitem{Bulbul:2014sua}
Esra Bulbul, Maxim Markevitch, Adam Foster, Randall~K. Smith, Michael
  Loewenstein, et~al.
\newblock {Detection of An Unidentified Emission Line in the Stacked X-ray
  spectrum of Galaxy Clusters}.
\newblock {\em Astrophys.J.}, 789:13, 2014, 1402.2301.

\bibitem{Boyarsky:2014jta}
Alexey Boyarsky, Oleg Ruchayskiy, Dmytro Iakubovskyi, and Jeroen Franse.
\newblock {An unidentified line in X-ray spectra of the Andromeda galaxy and
  Perseus galaxy cluster}.
\newblock 2014, 1402.4119.

\bibitem{Boyarsky:2014ska}
Alexey Boyarsky, Jeroen Franse, Dmytro Iakubovskyi, and Oleg Ruchayskiy.
\newblock {Checking the dark matter origin of 3.53~keV line with the Milky Way
  center}.
\newblock 2014, 1408.2503.

\bibitem{Jeltema:2014qfa}
Tesla~E. Jeltema and Stefano Profumo.
\newblock {Dark matter searches going bananas: the contribution of Potassium
  (and Chlorine) to the 3.5 keV line}.
\newblock 2014, 1408.1699.

\bibitem{Malyshev:2014xqa}
D.~Malyshev, A.~Neronov, and D.~Eckert.
\newblock {Constraints on 3.55 keV line emission from stacked observations of
  dwarf spheroidal galaxies}.
\newblock 2014, 1408.3531.

\bibitem{Boyarsky:2014paa}
A.~Boyarsky, J.~Franse, D.~Iakubovskyi, and O.~Ruchayskiy.
\newblock {Comment on the paper "Dark matter searches going bananas: the
  contribution of Potassium (and Chlorine) to the 3.5 keV line" by T. Jeltema
  and S. Profumo}.
\newblock 2014, 1408.4388.

\bibitem{Muhlleitner:2003me}
Margarete Muhlleitner and Michael Spira.
\newblock {A Note on doubly charged Higgs pair production at hadron colliders}.
\newblock {\em Phys.Rev.}, D68:117701, 2003, hep-ph/0305288.

\bibitem{Chatrchyan:2012vca}
Serguei Chatrchyan et~al.
\newblock {Search for a light charged Higgs boson in top quark decays in $pp$
  collisions at $\sqrt{s}=7$ TeV}.
\newblock {\em JHEP}, 1207:143, 2012, 1205.5736.

\bibitem{ATLAS-CONF-2013-090}
{Search for charged Higgs bosons in the $\tau$+jets final state with pp
  collision data recorded at $\sqrt s=8$ TeV with the ATLAS experiment}.
\newblock Technical Report ATLAS-CONF-2013-090, CERN, Geneva, Aug 2013.

\bibitem{DelNobile:2009st}
Eugenio Del~Nobile, Roberto Franceschini, Duccio Pappadopulo, and Alessandro
  Strumia.
\newblock {Minimal Matter at the Large Hadron Collider}.
\newblock {\em Nucl.Phys.}, B826:217--234, 2010, 0908.1567.

\bibitem{Aad:2014vma}
Georges Aad et~al.
\newblock {Search for direct production of charginos, neutralinos and sleptons
  in final states with two leptons and missing transverse momentum in $pp$
  collisions at $\sqrt{s} =$ 8 TeV with the ATLAS detector}.
\newblock {\em JHEP}, 1405:071, 2014, 1403.5294.

\bibitem{Khachatryan:2014qwa}
Vardan Khachatryan et~al.
\newblock {Searches for electroweak production of charginos, neutralinos, and
  sleptons decaying to leptons and W, Z, and Higgs bosons in pp collisions at 8
  TeV}.
\newblock 2014, 1405.7570.

\bibitem{Merle:2011yv}
Alexander Merle and Viviana Niro.
\newblock {Deriving Models for keV sterile Neutrino Dark Matter with the
  Froggatt-Nielsen mechanism}.
\newblock {\em JCAP}, 1107:023, 2011, 1105.5136.

\bibitem{Barry:2011fp}
James Barry, Werner Rodejohann, and He~Zhang.
\newblock {Sterile Neutrinos for Warm Dark Matter and the Reactor Anomaly in
  Flavor Symmetry Models}.
\newblock {\em JCAP}, 1201:052, 2012, 1110.6382.

\bibitem{Shaposhnikov:2006nn}
Mikhail Shaposhnikov.
\newblock {A Possible symmetry of the nuMSM}.
\newblock {\em Nucl.Phys.}, B763:49--59, 2007, hep-ph/0605047.

\bibitem{Lindner:2010wr}
Manfred Lindner, Alexander Merle, and Viviana Niro.
\newblock {Soft $L_e - L_\mu - L_\tau$ flavour symmetry breaking and sterile
  neutrino keV Dark Matter}.
\newblock {\em JCAP}, 1101:034, 2011, 1011.4950.

\bibitem{Merle:2013gea}
Alexander Merle.
\newblock {keV Neutrino Model Building}.
\newblock {\em Int.J.Mod.Phys.}, D22:1330020, 2013, 1302.2625.

\end{thebibliography}
\end{document}